\documentclass[aip,apl,a4paper,superscriptaddress,twocolumn,showpacs]{revtex4}
\usepackage{graphicx,graphics,color,epsfig}
\usepackage{amsmath}
\usepackage{amsfonts}
\usepackage{amssymb}

\begin{document}
\title{Negative differential resistances with back gate-controlled
lowest operation windows in graphene double barrier resonant tunneling diodes}
\author{Yu Song}
\email{kwungyusung@gmail.com}
\affiliation{Department of Physics and State Key Laboratory of Low-Dimensional
Quantum Physics, Tsinghua University, Beijing 100084, People's Republic of China}

\author{Han-Chun Wu}
\affiliation{School of Physics and CRANN, Trinity College Dublin, Dublin 2, Ireland}

\author{Yong Guo}
\affiliation{Department of Physics and State Key Laboratory of Low-Dimensional
Quantum Physics, Tsinghua University, Beijing 100084, People's Republic of China}

\begin{abstract}
We theoretically investigate negative differential resistance (NDR)
of massless and massive Dirac Fermions in double barrier resonant
tunneling diodes based on sufficiently short and wide graphene strips. The
current-voltage characteristics calculated in a rotated pseudospin
space show that, the NDR feature only presents with appropriate
structural parameters for the massless case and the peak-to-valley
current ratio can be enhanced exponentially by a tunable band gap.
Remarkably, the lowest NDR operation window is nearly structure-free
and can be almost solely controlled by a back gate, which may have potential applications
in NDR devices with the operation window as a crucial parameter.
\end{abstract}
\date{\today}
\maketitle


Negative differential resistance (NDR) is a fundamental physical phenomenon
which has been observed in various systems, including
gaseous media,\cite{gas}
chalcogenide glasses,\cite{glass}
organic semiconductors,\cite{organic}
conductive polymers,\cite{polymer} etc.
In electronics, NDR is widely used in high-speed applications
including high-frequency signal generation and high-speed switching,
and functional applications such as one-transistor static memories
and multi-valued memory circuits.\cite{oscillator}
Recently, extensive efforts\cite{SBD,zigzag,armchair,junction,SL,classical}
have been devoted to the study of NDR in graphene, a monolayer of $sp^2$ bonded carbon atoms that has attracted much
attentions since its discovery.\cite{graphene_realize}
NDR features in graphene single barrier diodes,\cite{SBD}
zigzag nanoribbons,\cite{zigzag}
armchair nanoribbons,\cite{armchair}
numerus nanoribbon junctions,\cite{junction}
armchair superlattices,\cite{SL} and three terminal field-effect
transistors\cite{classical} have been theoretically or experimentally reported.

It is well-known that resonant tunneling (or equivalently Fabry-P\'{e}rot-type
interference\cite{FP}) is a fundamental mechanism for NDR;\cite{oscillator}
it plays a dominant role in the NDR feature in common semiconductor
based resonant tunneling diodes (RTDs).\cite{oscillator,NDR}
Surprisingly, so far this basic mechanism has not been explored in graphene except
the armchair superlattices work.\cite{SL}
However, in this structure, other mechanisms (band gaps, miniband conductance, and Wannier-Stark ladder)
also contribute to the NDR feature,\cite{SL}
thus significantly obscure the resonant tunneling mechanism.

\begin{figure}[h]
\centering
\includegraphics[width=0.9\linewidth]{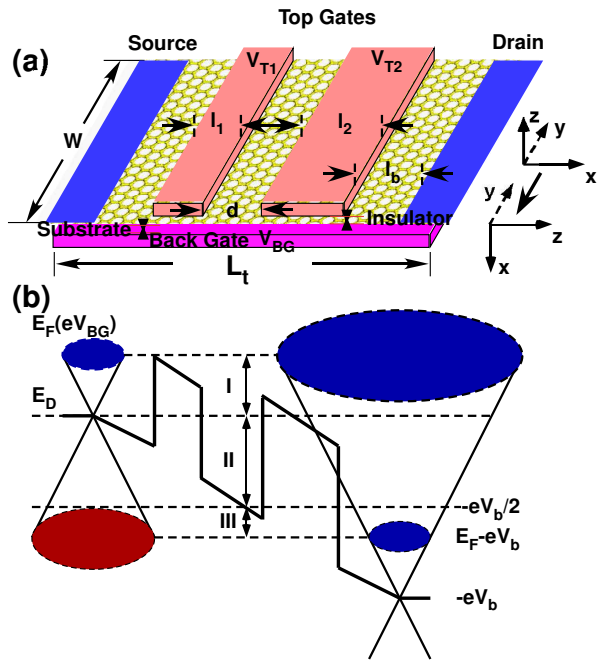}
\caption{(color online)
Model construction for I-V characteristics of graphene DB RTDs.
(a) The diode contains two barriers (with lengths $l_{1(2)}$
and height $V_{t1(2)}$) separated by a well (with length $d$),
and two buffer regions (with length $l_b$) separating the barriers and the electrodes.
The two rectangular coordinates show the rotation we make in the pseudospin space.
(b) The biased transport can be divided into three regimes defined by
the Fermi energy $E_F$ and the finite bias $V_b$.
In these regimes, electron-to-electron (I), hole-to-electron (II), and
restricted hole-to-electron transport (III) respectively contributes to the DC.
In regime I (II) the bias induced DC is approximately
proportional to $E-eV_b$ ($eV_b-E$) (see, Ref. 12), 
while in regime III 
only holes within critical incident angles $\mp\sin^{-1}(1+eV_b/E)$ contribute to the
transport (that's why we call it restricted).}
\end{figure}

In this letter, we theoretically investigate the NDR feature of massless
and massive Dirac Fermions in double barrier (DB) RTDs based on sufficiently short
and wide graphene strips (see, Fig. 1(a)).
We consider a realistic linear voltage drop between the source and drain electrodes (see, Fig. 1(b)) and
calculate the current-voltage characteristics in a rotated pseudospin
space (see, Fig. 1(a)) within the Landauer-B\"{u}ttiker formalism.
We find that the NDR only appears with appropriate structural parameters
for the massless case and the peak-to-valley current ratio can be
enhanced nearly exponentially by a tunable band gap (i.e., the mass term).
Remarkably, we also find that the lowest NDR operation window (the
bias range between the current peak and valley) is nearly free to the
structural parameters and is always locked
around the Fermi energy hence can be almost solely controlled by a back gate.
This phenomenon could be of benefit to NDR devices
in which the operation window plays a dominant role.\cite{oscillator}

The structure of the graphene DB RTDs is shown in Fig. 1(a).
A graphene strip with a dimension of $L_t\times W$ is placed on a substrate 
in the $x$-$y$ plane.
Here $W$ is several times of $L_t$ to ensure that the edge effect is negligible.\cite{poisson}
The graphene strip is further contacted by a source and drain
electrode along the $y$-direction and isolated
by an insulator layer on top of it. 
When made of high-$\kappa$ (dielectric constant) material,\cite{high-k}
the contact can be regarded as ideal, i.e., the contact-induced energy
broadening and a finite contact resistance\cite{contact} can be ignored.
The DB RTD can be fabricated by patterning
two top gates\cite{exper} ($V_{t1}$ and $V_{t2}$) on top of the insulator
layer along the $y$-direction, and contacting a back gate ($V_{BG}$) to the substrate. 
The realistic barriers formed by the top gates are smooth due to the interface electric field.\cite{width}
However, they can be regarded as rectangular ones with the same lengths but effective
heights determined by the smoothness of the realistic barriers.\cite{giantGH}
The carrier concentration ($n$) in the graphene strip is linearly tuned
by the back gate.\cite{graphene_realize}
Accordingly the Fermi energy (respective to the graphene charge neutrality
point, i.e., the Dirac point $E_D\equiv0$)
is also tuned by the back gate since $E_F\propto \textmd{sign}(n)\sqrt{|n|}$.\cite{fermi}
When a bias voltage ($V_{b}$) is applied between the source and drain,
a linear voltage drop along the $x$-direction will be formed due to a uniform in-plane electric
field (see, Fig. 1(b)).
Meanwhile, a net current will be produced by
the electrons or holes in the source electrode within an energy range of $E\in[E_{F}-eV_{b}, E_{F}]$ at zero-temperature.


To calculate the I-V characteristics in the Landauer-B\"{u}ttiker
formalism,\cite{LB} one need to first solve envelope function ($\boldsymbol{\Psi}(x,y)=(\psi^\uparrow,\psi^\downarrow)^T$
where $\uparrow/\downarrow$ corresponds to the $A$/$B$ sublattice)
in each uniform region.
Straightforward decouple of the original Dirac equation containing
the electric field along the $x$-direction, however, unfortunately results in an
unsolvable two-order differential equation.
Here we perform a rotation of the Dirac equation by $\pi/2$ around
the $y-$axis in the \emph{pseudospin} space (see, Fig. 1(a)).
The Hamiltonian becomes\cite{rotation}
\begin{equation}
H=v_{F}(\sigma_{z}p_{x}+\sigma_{y}p_{y}-\sigma_x\Delta)+eV(x)\mathbf{I}.
\end{equation}
Here $v_F\approx10^6m/s$ is the Fermi velocity, $\boldsymbol{\sigma}=(\sigma_x,\sigma_y,\sigma_z)$
are Pauli's matrices,
$\mathbf{p}=(\hbar k_x,\hbar k_y)^T$ is the momentum operator, $\Delta$ is a tunable
band gap (i.e., the mass of the Dirac Fermions) up to several hundreds of meV achieved through a controllable
doping,\cite{gap} $V(x)$ is the position dependent electrostatic potential,
and $\mathbf{I}$ is the 2$\times$2 identity matrix.
For convenience we express all the parameters in their dimensionless form:
$x=x/l_0$, $k=k l_0$, $\epsilon=E/E_0$, $\delta=\Delta/E_0$, and $v(x)=eV(x)/E_0$
in terms of a characteristic length $l_0$ and corresponding energy unit $E_0\equiv\hbar v_F/l_0$.
$l_0$ is adopted as 40 nm ($E_0\approx16.44$ meV)
to ensure the electron density of states coinciding with a true system
and a coherent transport regime even at room temperature.

The envelope functions in the buffer and well (barrier) regions can be
exactly solved from the decoupled two-order differential equation
\begin{equation}
\boldsymbol{\Psi}=p\left(\begin{array}{c}
F\\
G\end{array}\right)e^{ik_{y}y}+q\left(\begin{array}{c}
G^{*}\\
F^{*}\end{array}\right)e^{ik_{y}y},
\end{equation}
where $F=D[-1+iq^2/2a,(1+i)(\epsilon+ax(-v_t))/\sqrt{a}]$ and
$G=(1+i)\sqrt{a}q^{-1}D[iq^2/2a,(1+i)(\epsilon+ax(-v_t))/\sqrt{a}]$
with $D$ being the Weber parabolic cylinder function, $q^2=k_y^2+\delta^2$, and
$a=eV_b/L$ ($L=2l_b+l_1+d+l_2<L_t$ is the total length between the source and drain).
Note, $F$ and $G$ ($F^{*}$ and $G^{*}$) have properties of a right
(left)-going wave function.\cite{rotation}
The decoupled two-order differential equation 
in the electrode regions recovers the one before performing rotation.
A proper envelope function should be adopted as $\boldsymbol{\tilde{\Psi}}(x,y)=(\psi^+,\psi^-)^T$
with the spinor components relating with the original ones 
by $\psi^{\pm}=(\pm\psi^{\uparrow}+\psi^{\downarrow})/\sqrt{2}$.

\begin{figure}[h]
\centering
\includegraphics[width=\linewidth]{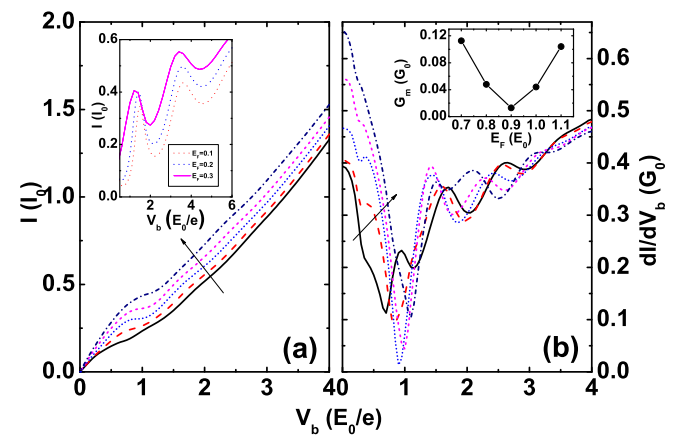}
\caption{(color online) (a) I-V and (b) DC characteristics for a graphene DB RTD with parameters of
$l_b=1$, $l_{1(2)}=1$, $d=5$, and $\upsilon_{t1(2)}=1$. Along the arrow, $\epsilon_{F}$=0.7, 0.8, 0.9, 1.0, and 1.1.
Insert in (a): I-V characteristics for a 2DEG based DB RTD with the same parameters
(note now $E_0=\hbar^2/2ml_0^2$ with $m$ being the effective electron mass in 2DEG).
Insert in (b): the minimum DC as a function of the Fermi energy.
}
\end{figure}



The transmission coefficient ($t$) can be obtained by matching the
spinor envelope functions at the potential boundaries with the standard
transfer-matrix method.\cite{TMM}
The transmission probability reads $T=k_{Dx}(\epsilon+\delta)k_{Sx}^{-1}(\epsilon+
\delta+v_{b})^{-1}|t|^{2}$ for $k_{Sx}^2>0$ and $k_{Dx}^2>0$,
and $T=0$ otherwise, where $k_{S(D)x}$ is the value of $k_x$ at the
source (drain) electrode. Then the net current at zero temperature can be calculated by the
Landauer-B\"{u}ttiker formalism\cite{LB}
\begin{equation}
I(V_{b})=I_0\int_{\epsilon_{F}-v_{b}}^{\epsilon_{F}}\int_{-\pi/2}^{\pi/2}T(\epsilon,\theta,v_{b})|\epsilon|\cos\theta d\theta d\epsilon,
\end{equation}
where $I_0=4ev_{F}W/(2\pi l_0)^2$ is a current unit with
the factor 4 coming from the spin and valley degeneracies.
Calculated current based on Eq. (2) avoids possible
nonphysical current induced by a steplike approximation of the voltage drop.\cite{SBD,SL} 


\begin{figure}[h]
\centering
\includegraphics[width=\linewidth]{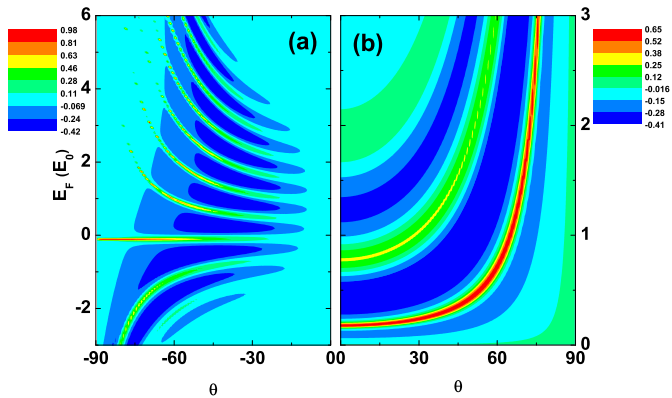}
\caption{(color online) Differences of transmission probabilities
between the unbiased DB RTD calculated in Fig. 2
and its constituted single barrier based on (a) gapless graphene
and (b) 2DEG. For both the graphene and 2DEG cases the differences are the same for $\pm \alpha$.
}
\end{figure}

Fig. 2(a) shows the I-V characteristics of a graphene DB RTD 
at various Fermi energies controlled by the back gate.
One can see that, the I-V curves display obvious ripples with a possible
NDR feature around $eV_b=E_0$.
To analyze whether there is a NDR or not, we further plot the differential
conductance (DC) in Fig. 2(b).
As is seen, when the bias sweeps from zero, the DC first decreases
and then increases successively with a relatively big and small
gradient. This can be understood by the three transport
regimes marked and described in Fig. 1(b).
Moreover, one can see clearly the oscillation behavior in DC, which is a result of the alternate
enhancement and suppression of $T$ respectively
at and between resonant tunneling peaks (see, Fig. 3(a)).
Note, for a given $E_F$ the DC achieves the minimum ($G_m$) at some bias.
We summary $G_m$ as a function of the Fermi energy in
the inset of Fig. 2 (b). One can see that it
first decreases and then increases with increasing $E_F$.
The global minimum DC for the considered structure appears at $\epsilon_F = 0.9$
with a value of about 0.013, which confirms that
there is no NDR in such a graphene DB RTD.

This is an interesting result comparing with the rather obvious NDR in two-dimensional
electron gas (2DEG) based DB RTD with the same structural parameters (see inset in Fig. 2(a)).
In this type of RTD's, NDR occurs when
the suppression regions of $T$ 
(i.e., blue regions in Fig. 3(b)) enter the integration window
of the current.\cite{NDR}
Interestingly, in graphene these suppression regions are significantly
reduced (see, Fig. 3(a)), especially for relatively small incident angles
which unfortunately make the main contribution to the NDR (see the factor $\cos\theta$ in Eq. (3)).
This is because the quasibound states (equivalently, resonant tunneling peaks)
are hard to form due to the Klein tunneling\cite{klein} in these regions.
Moreover, the integration window in graphene is $[E_F-eV_b,E_F]$ rather
than $[\textmd{Max}(0, E_F-eV_b),E_F]$ in 2DEG or common semiconductors.
Here, $\textmd{Max}(u,v)$ stands for the bigger one of $u$ and $v$.
Then when $eV_b$ exceeds $E_F$, the hole-to-electron transport in transport regimes II and III
(which is absent in the semiconductors case) also contributes a positive DC in
the I-V curves of graphene. This DC increases with
increasing bias hence further suppresses the NDR feature in graphene.
Therefore, the I-V characteristics and NDR features in graphene DB RTDs
are a competition of hole-to-electron transport and Klein tunneling with the
resonant tunnelings.




\begin{figure}[h]
\centering
\includegraphics[width=\linewidth]{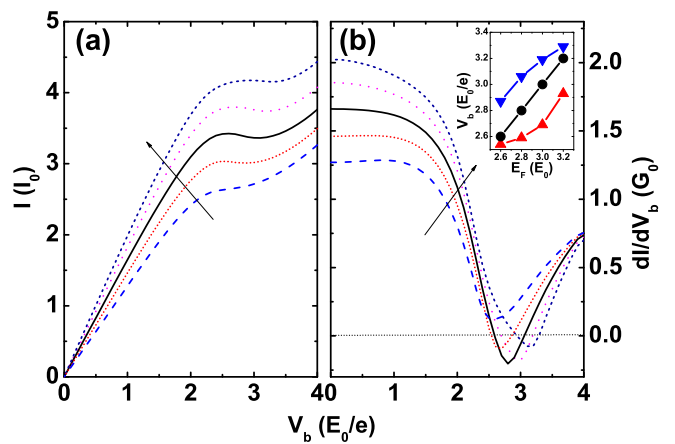}
\caption{(color online) (a) I-V and (b) DC characteristics for a
graphene DB RTD with parameters of $l_b=1/2$, $l_{1(2)}=1/2$, $d=1/2$,
and $\upsilon_{t1(2)}=2$. Along the arrow, $\epsilon_{F}$=2.4, 2.6,
2.8, 3.0, and 3.2. Insert in (b): the bias positions for the current
peak (curve with $\blacktriangle$), valley ($\blacktriangledown$),
and minimum DC ($\bullet$) as a function of the Fermi energy.
}
\end{figure}

The absence of NDR can be overcome by enhancing the resonant tunneling
in DB RTDs with more appropriate structural parameters.
We find that the less the quasibound states (which
approximately equals to the value of $v_td$), the stronger the
contribution of the resonant tunneling.
Fig. 4 shows the I-V characteristics and DC's for a DB RTD with
$l_b=1/2$, $l_{1,2}=1/2$, $d=1/2$, and $v_{t1,2}=2$ at different
Fermi energies.
As is seen, $G_m$ becomes negative, i.e., the NDR is obtained
for $\epsilon_F\approx2.6-3.2$.
It is very interesting to note that, $G_m$ is always locked exactly
at the bias of $eV_b=E_F$ (see, the insert in Fig. 4(b)).
This feature is essentially different from the case of common
semiconductors based DB RTDs, and has not been found in other
types of graphene NDR structures.
In semiconductors based DB RTDs, the bias positions for the current peaks are
determined by the excited state levels of the quantum well.\cite{NDR}
Then the operation windows (OW's) are almost Fermi energy-free and synthetically
controlled by the structural parameters (i.e., $l_{1,2}$, $v_{t1,2}$, and $d$).
In contrast, for graphene DB RTDs,
the central position for the lowest OW is almost structure-free and depends only on the Fermi energy.
Then by solely tuning the back gate voltage the Fermi energy and hence the
lowest OW can be exactly controlled or chosen as long as the NDR is present.

It is found in Fig. 2(b) that, the biases for local minimum DC's generally
decrease with increasing Fermi energy (the one around
$v_b=1.2$ for $\epsilon_F=0.7$ is an exception).
Note, for $G_m$ the bias increases almost linearly as a function of $E_F$.
On the other hand, for $eV_b<E_F$ it is purely electron-to-electron transport,
while for $eV_b>E_F$ both electron-to-electron and hole-to-electron
transports make contribution.
So, the ambipolar transport is at the heart of the physics for the
remarkable back gate-controlled lowest OW.
Note the lower output voltage in high-speed switching circuits and
the two stored states in static memory elements are determined
exactly by the OW (see Ref. 5 
and relevant references therein).
Such back gate-controlled OW would have potential
applications in these NDR based devices.


\begin{figure}[h]
\centering
\includegraphics[width=\linewidth]{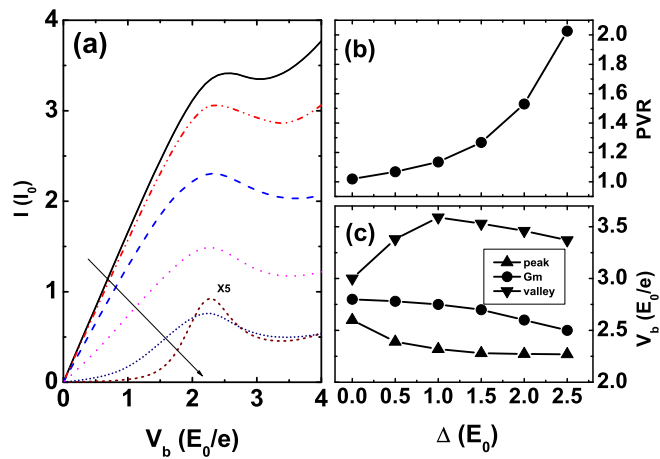}
\caption{(color online) (a) I-V characteristics for the graphene DB RTD
considered in Fig. 4 ($\epsilon_F=2.8$) with different band gaps.
Along the arrow, $\delta$=0, 0.5, 1.0, 1.5, 2.0, and 2.5 (enlarged by 5 times for clearness).
(b) The peak-to-valley current ratio and (c) the biases for the current
peaks, $G_m$, and current valleys as a function of the band gap.
}
\end{figure}

The PVR is still rather small (about 1.02 for the strongest
case $\epsilon_F=2.8$) for these gapless graphene based DB RTDs.
In Fig. 5(a) we show the I-V characteristics for graphene DB RTDs with
different band gaps.
It is clear that, the current at a fixed bias decreases as
the band gap (which should not exceed the Fermi energy) increases.
When the band gap is sufficiently large (i.e., $\delta\geq1.5$), the low bias I-V
characteristics  becomes superlinear and the second
NDR OW with a much smaller PVR appears at a higher bias.
Moreover, the PVR for the lowest OW increases almost exponentially with
increasing band gap ($\textrm{PVR}\approx0.9832+0.0413\textrm{e}^{1.29\delta}$ for $\delta<2.8$)
(see, Fig. 5(b)).
The underlying physics for such drastic enhancement is that, the
presence of band gap not only suppresses the hole-to-electron transport
and Klein tunneling (adverse factors to
the NDR), but also
enhances the resonant tunneling hence the reduction region of $T$ (a
favorable factor to the NDR),
as the modulus of the image longitudinal
wave vector in the barriers ($\kappa_x(x)=[\delta^2+k_y^2-(\epsilon-v(x))^2]^{1/2}$) increases with the band gap.

In Fig. 5(c) one can see that, the width of the lowest NDR OW first
increases and then slightly decreases with the band gap. However, its
central position is nearly independent on the band gap
thus also can be almost solely controlled by the back gate.
Our investigations further show that, proper structural asymmetries with
the right barrier higher and/or longer than the left barrier can slightly
enhance the PVR as optimal resonant tunneling happens under bias.\cite{optimal}
However, the relation of $eV_b=E_F$ for $G_m$ is broken.
Then graphene symmetric DB RTDs should be adopted for utilizing the
back gate-controlled lowest OW.

Very recently, electrostatic junctions and hence Fabry-P\'{e}rot-type
interferences induced by the source and drain metal contacts are experimentally
reported.\cite{metalcontact}
Such junctions can be modeled by a positive ($p$ doping) electrostatic
potential $V_{MC}$ for the graphene underneath the metal and
a Fermi energy $E_F$ (which can be tuned by the back gate) through the whole structure.\cite{metalcontact}
Then the source and drain possess an effective Fermi energy $E_F-eV_{MC}$ and
two extra quantum wells form between the two ports and the DB region.
As a result, the I-V characteristics will quantitatively
change especially in the bias range $eV_b\in(E_F-eV_{MC},E_F)$,
where the integral windows overlap with the energy range $(0,eV_{MC})$
of the extra quantum wells.
However, since the physical mechanism do not change, the qualitative trends of the I-V
characteristics for both the gapless and gapped cases would be the same as the $V_{MC}=0$ cases discussed above.
As the transport is still ambipolar, the lowest OW can also be
almost solely controlled by the back gate. Due to the shifted source and drain
effective Fermi energy, it will be found around
a shifted bias $eV_b=E_F-eV_{MC}$.



In summary, we have theoretically investigated the NDR in graphene
symmetric DB RTDs and demonstrated an almost structure-free and
back gate-controlled lowest OW. This remarkable phenomenon stems
from the ambipolar transport in graphene and may be applied in OW-dominated NDR devices.
We have also found that, appropriate structural parameters are necessary for the
NDR feature and a tunable band gap can enhance exponentially the PVR.
The competition between hole-to-electron transport, Klein tunneling,
and resonant tunneling is the main mechanism for such a NDR structure.


This work was supported by the NSFC (10974109 and 11174168),
the SRFDP (20100002110079), and the 973 Program of China (2011CB606405).

\end{document}